\documentclass{ws-ijgmmp-web}
\usepackage{color}
\usepackage[colorlinks=true]{hyperref} 

\newcommand\tr{\operatorname{tr }}

\def\e{\mathrm{e}}
\def\ii{\mathrm{i}}
\def\d{\mathrm{d}}

\begin{document}

\markboth{P. Facchi, G. Garnero, M. Ligab\`o}
{Quantum Thermodynamics and Canonical Typicality}
%
\catchline{}{}{}{}{}
%

\title{QUANTUM FLUCTUATION RELATIONS}

\author{PAOLO FACCHI, GIANCARLO GARNERO} 
\address{Dipartimento di Fisica and MECENAS,
Universit\`a di Bari, I-70126 Bari, Italy\\ 
INFN, Sezione di Bari,  I-70126 Bari, Italy}

\author{MARILENA LIGAB\`O}
\address{Dipartimento di Matematica,
Universit\`a di Bari, I-70125 Bari, Italy}

\maketitle

\begin{abstract}
We present here a set of lecture notes on exact fluctuation relations. We prove the Jarzynski equality and the Crooks fluctuation theorem, two  paradigmatic examples of classical fluctuation relations. Finally we consider their quantum versions, and analyze analogies and differences with the classical case.
\end{abstract}

\keywords{quantum thermodynamics; fluctuation relations.}

\setcounter{section}{-1}
\section{Introduction} 

\label{sec:intro}

\vspace{0.3cm}

We present here the notes of three lectures given by one of us at the  ``Fifth International Workshop on Mathematical Foundations of Quantum Mechanics and its applications" held in February 2017 in Madrid at the \textit{Instituto de Ciencias Matem\'aticas} (ICMAT).

We will consider some results about fluctuation theorems both for classical and for quantum systems,  a research topic that recently has attracted a great deal of attention. The statistical mechanics of classical and quantum systems driven far from equilibrium has witnessed quite  recently a sudden development with the discovery of various exact fluctuation theorems which connect equilibrium thermodynamic quantities to non-equilibrium ones. There are excellent reviews on this topic, which cover both classical~\cite{seifert} and quantum fluctuation relations~\cite{esposito,campisi}. Here we will follow more closely the exposition by Campisi, H\"anggi and Talkner~\cite{campisi}, to which we refer the reader for further information.

In the first lecture we will  recall the derivation of Einstein's fluctuation-dissipation relation for a Brownian particle, which is the inception of classical fluctuation relations. Moreover, we will  identify the fundamental ingredients which are already present in this early derivation.
Then, we will consider the Green-Kubo formula, which represents the first general approach  to quantum fluctuation-dissipation relations. The link between the correlation function of a quantum system and its linear response function will be shown and  the classical limit of the Green-Kubo formula will be considered.

The second lecture is devoted to  exact classical fluctuation relations. In particular, we will give  explicit deductions of the Jarzynski equality  and the Crooks fluctuation theorem, two paradigmatic examples of classical fluctuation theorems. In the proofs two ingredients will be crucial: reversibility at the microscopic level and the Gibbs probability distribution on the initial conditions of the system. 

Finally, in the third lecture we will consider the quantum case. After introducing the operational definition of measurement of work as a two-time energy measurement, and properly defining  microreversibility for time-dependent unitary evolution, we prove both the Jarzynski equality and  the Crooks fluctuation theorem for a quantum system.

\section{Lecture 1: Fluctuation-dissipation relations}
\label{sec:lecture1}

We start with some classical results about fluctuation-dissipation relations.

At the microscopic level matter is in a permanent state of agitation and undergoes thermal and quantum fluctuations. Statistical Mechanics is able to provide explanations and quantitative results on those fluctuating quantities.

A paradigmatic example is a rarified gas at thermal equilibrium, which is classically described by the Maxwell-Boltzmann distribution of velocities. This distribution is derived under the assumption that the classical dynamics of the microscopic constituents is Hamiltionian, and that the atoms of the gas interact via negligible short-range forces.
Moreover, the Maxwell-Boltzmann describes a situation of thermal equilibrium. 

What happens to other fluctuating quantities? 

In this lecture we will be mainly interested in the work exchanged  during out-of-equilibrium  transformations and its  fluctuations.
In this analysis a crucial role will be played by two ingredients:
\begin{itemize}
\item the initial state of a physical system is  a thermal state, and, as such, is described by the Gibbs canonical distribution:
\begin{equation}
\rho_\beta=\frac{{\rm e}^{-\beta H_0}}{Z_0}
\end{equation}
where $\beta^{-1}\propto T$ is the temperature at equilibrium, $H_0$ is the Hamiltonian at the initial time, and $Z_0$ is the partition function; 
\item the dynamics is reversible at the microscopic level.
\end{itemize}
The first hypothesis is of statistical nature, because we are assuming a well defined initial probability distribution on the initial state. On the other hand, the second one is only stating the Hamiltonian nature of microscopic dynamics.

Next, we would like to understand what happens after forcing the system out of equilibrium, not necessarily in an adiabatic way.
\begin{figure}[tbp]
\label{brownian}
\centering
\includegraphics[width=0.6\columnwidth]{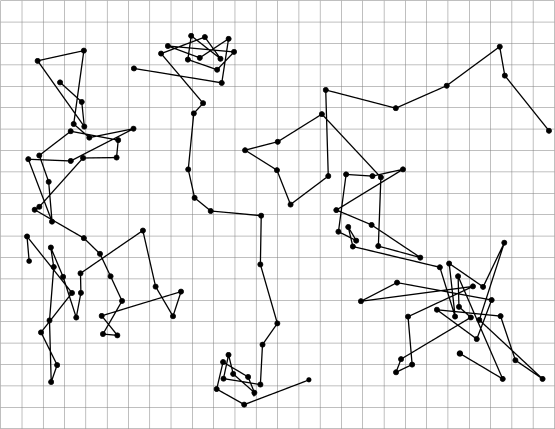}
\caption{Evidence of Brownian motion as depicted for the first time by Jean Perrin in 1908~{\protect\cite{perrin}}.}
\end{figure}

\subsection{Einstein's relation}

The history of fluctuation relations can be traced back to the work of Einstein \cite{einstein}.
In 1905 he proved that the linear response of a system in thermal equilibrium, driven out of equilibrium by an external force, is determined by the fluctuation properties at equilibrium.

Einstein considers the case of a Brownian particle in a fluid (see Figure \ref{brownian}) and determines a relation between the mobility $\mu$ and the diffusion constant $D$:
\begin{equation}
\label{eq:1905}
\boxed{
\mu=\frac{D}{k_B T},
}
\end{equation}
where $k_B$ is the Boltzmann constant and $T$ is the absolute temperature of the fluid at equilibrium.
We recall that, in a dissipative fluid, the mobility represents the ratio of the suspended particle's terminal drift velocity $v_d$ to an applied force $F$:
\begin{equation}
\mu=\frac{v_d}{F},
\end{equation}
It is apparent from equation~\eqref{eq:1905} that Einstein's relation links a non-equilibrium quantity, say $\mu$, related to the force that drags the system out of its initial state, with  the temperature $T$ of the gas at equilibrium.

We briefly recall the derivation of Einstein's relation. Suppose the force $F$ is conservative, say $F(x)=-\nabla U(x)$, where $U:\mathbb{R}^3 \to \mathbb{R}$ is a smooth potential. Then, the drift velocity at $x$ reads 
\begin{equation}
v_d(x)=\mu(x)F(x)=-\mu(x) \nabla U(x).
\end{equation}
Moreover, assume that the concentration is at equilibrium and thus is determined by the Maxwell-Boltzmann statistics,
\begin{equation}
\rho(x)= A\,  \e^{-\frac{U(x)}{k_B T}},
\label{eq:MB}
\end{equation} 
where $A$ is a normalization constant. 
The  current density due to drift reads 
\begin{equation}
J_{\rm drift}(x)=\rho(x) v_d (x)=-\rho (x) \mu(x) \nabla U(x). 
\end{equation}
There is a second contribution to the current density  which is due to diffusion and, according to Fick's law, is proportional to the gradient of the concentration:
\begin{equation}
J_{\mathrm{diffusion}}(x)=-D(x) \nabla \rho(x).
\end{equation}
At  equilibrium there is a  balance between these currents, namely
\begin{equation}
\label{eq:balance}
J_{\rm drift}(x)+J_{\rm diffusion}(x)=0.
\end{equation} 
By deriving~\eqref{eq:MB} we obtain
\begin{equation}
\nabla \rho(x)=-\frac{\nabla U(x)}{k_B T}\rho(x)
\end{equation}
and plugging it in the balance equation~\eqref{eq:balance} we get
\begin{equation}
-\rho(x) \nabla U(x)\left[\mu(x) - \frac{D (x)}{k_B T}\right]=0,
\end{equation}
and Einstein's fluctuation-dissipation relation~\eqref{eq:1905}  follows. It is evident from the above derivation that~\eqref{eq:1905} is an approximate relation, since  Fick's law is only valid in the linear regime. 

Einstein's relation was the first of a series of fluctuation-dissipation relations, which  predict the behavior of systems that obey the detailed balance principle and are weakly perturbed from thermal equilibrium:  thermal fluctuations of a physical observable are related to  the linear response, quantified by the admittance or impedence of the same physical observable.
The key idea is that the response of a system, which is at thermodynamic equilibrium, to a small applied force is the same as the response to statistical fluctuations at equilibrium.

A second example of a fluctuation-dissipation relation was provided by the Johnson-Nyquist noise~\cite{john,nyq}. This  phenomenon is due to the thermal agitation of electrons in a conductor at  equilibrium. The overall effect is an electrical thermal  noise which can be measured and appears as a difference voltage acting at the extrema of an isolated resistor.  This time-dependent voltage, known as noise voltage, depends on the conductor's temperature and its mean square value is given by
\begin{equation}
\langle V^2\rangle=4 R\, \Delta \nu\, k_B T,
\end{equation}
where $R$ is the resistance and $\Delta \nu$ is the bandwidth of the observed frequencies. 

\subsection {Green-Kubo relations}
Let us now quickly review the general framework of the fluctuation-dissipation relations provided in a quantum-mechanical setting by the Green-Kubo relations~\cite{green, kubo}.

Consider and isolated quantum system, whose  Hamiltonian operator is $H_0$, which is a self-adjoint operator on a Hilbert space $\mathcal{H}$. Suppose that the system is at thermal equilibrium at temperature $T$, say
\begin{equation}
\label{eq:gibbs}
\rho_\beta=\frac{ \e^{-\beta H_0}}{Z_0},
\end{equation}
where $\beta=1/k_B T$,  and $Z_0={\rm Tr} ( \e^{-\beta H_0})$ is the partition function.
Assume that the system is perturbed  by an external time-dependent force, so that the total Hamiltonian reads
\begin{equation} \label{eqn:hamilt}
H(\Lambda_t)= H_0-\Lambda_t Q,
\end{equation}
where $t \in [0,\tau] \mapsto \Lambda_t \in \mathbb{R}$, with $\tau >0$, and $Q=Q^\dagger$ is the observable  coupled to the force $\Lambda_t$. For simplicity we shall assume that $Q$ is bounded.

The motion of the system is perturbed by the force $\Lambda$, but the perturbation is small if the force is weak. We will confine ourselves to weak perturbations and look at the response of the system in the linear approximation. The response is observed through the average change $\Delta B(t)$ of the bounded observable $B$. It is not difficult to prove~\cite{kubo} that, at the first order in $\Lambda$,
\begin{equation}
 \Delta B(t) :={\rm Tr} (B \rho_t)-{\rm Tr} (B\rho_\beta)=\int_0^t\Phi_{BQ}(t-s)\Lambda_s \mathrm{d}s,
\end{equation}
where $\rho_t$ denotes the evolution at time $t$ of $\rho_\beta$ under the action of the perturbed Hamiltonian~(\ref{eqn:hamilt}).
The kernel $\Phi_{BQ}$ is the so called response function and is given by
\begin{equation}
\Phi_{BQ}(t)=\frac{\langle [Q,B(t)] \rangle_\beta}{i\hbar},
\end{equation}
where  $[A,B]=AB-BA$ is the commutator, 
\begin{equation}
B(t)= \e^{\ii H_0 t/\hbar}B \e^{-\ii H_0 t/\hbar} \qquad (t\in \mathbb{R})
\end{equation}
is the (unperturbed) evolution of the observable $B$, and $\langle A \rangle_\beta = \tr (A \rho_\beta)$ is the thermal expectation value.

The second ingredient is the correlation function $\Psi_{BQ}$:
\begin{equation}
\Psi_{BQ}(t)=\frac{\langle \{Q,B(t)\} \rangle_\beta}{2},
\end{equation}
where $\{A,B\}=AB+BA$ is the  anticommutator.

The quantum fluctuation-dissipation theorem \cite{callen} links the above two functions, that is
\begin{equation}
\boxed{
\hat{\Psi}_{BQ}(\omega)=\frac{\hbar \omega}{2}\mathrm{coth} \left(\frac{\beta \hbar\omega}{2}\right)\hat{\Phi}_{BQ}(\omega),}
\end{equation}
where $\hat{f}$ denotes the Fourier transform of the function $f$,
\begin{equation}
\hat{f}(\omega)=\int_{\mathbb{R}} \e^{-\ii\omega t} f(t)\mathrm{d}t.
\end{equation}
Notice that the classical limit, $\hbar\to0$, of the quantum fluctuation-dissipation theorem reads
\begin{equation}
\hat{\Psi}(\omega)=\beta \hat{\Phi}(\omega),
\end{equation}
since $\mathrm{coth}(x)\sim 1/x$ as $x\to0$. The classical limit is in accordance with  Einstein's relation \eqref{eq:1905}.
The Green-Kubo relations started a new trend of research on higher order fluctuation-dissipation relations beyond the linear regime~\cite{evans,gallavotti,jarzynski}.

\section{Lecture 2: Classical fluctuation relations}
In this section we are going to deduce the so-called \emph{Jarzynksi equality}~\cite{jarzynski} and the \emph{Crooks fluctuation theorem}~\cite{crooks}, which are two paradigmatic examples of exact classical fluctuation theorems.

Consider a fluctuating quantity $x$ (for example the number of transported electrons in a resistance, or the heat, or the work in non-equilibrium transformations) and call $p_F(x)$ (the subscript $F$ stands for \emph{forward}) the probability density function of $x$ during a non-equilibrium thermodynamic transformation; call $p_B(x)$ the probability density function of the same quantity $x$ but under the backward transformation (the subscript $B$ stands for \emph{backward}). 
Then, due to microreversibility, fluctuation relations are usually expressed as a link between $p_F(x)$ and $p_B(x)$ of the form
\begin{equation}
\label{eq:fluct}
p_F(x)=   \e^{ \beta(x-a)} p_B(-x),
\end{equation}
where $a$ is a quantity related to the equilibrium starting points of the forward and backward processes.

Equation~\eqref{eq:fluct} relates non-equilibrium quantities, say the probability distributions of $x$ for the forward and backward processes, to equilibrium quantities, say the constants $\beta$ and $a$.
Thus, microreversibility implies that at the macroscopic level, the forward probability is exponentially more likely than the backward one. For example it could happen that the entropy of a small isolated system might spontaneously decrease, e.g. the water in a glass could spontaneously freeze. However, the relation~\eqref{eq:fluct} mantains that this process is extremely highly unprobable.

As already discussed in Lecture~\ref{sec:lecture1}, in the analysis of classical fluctuation relations two ingredients are fundamental:
\begin{itemize}
\item  reversibility at microscopic scales;
\item  initial  condition at thermal equilibrium.
\end{itemize}

We would like to analyze the work fluctuations for a classical non-autonomous system and deduce, as a consequence, the corresponding exact fluctuation relation.
Consider a classical system described by the Hamiltonian function
\begin{equation}\label{eq:clham}
H(z,\lambda)=H_0(z)-\lambda Q(z),
\end{equation}
where $z= (q,p)$ is a point in the phase space $\Gamma$ of the physical system, $H_0$ is the unperturbed Hamiltonian function, and  $\lambda$ is a real  parameter representing the external force coupled to the conjugate variable $Q(z)$. We assume that both $H_0$ and $Q$ are smooth functions on $\Gamma$.

The Gibbs canonical state at temperature $\beta^{-1}$ associated to the Hamiltonian~\eqref{eq:clham} is
\begin{eqnarray}
\rho_\beta^\lambda (z)&=& \frac{ \e^{-\beta H(z,\lambda)}}{Z(\lambda)}=\frac{ \e^{-\beta (H_0(z)-\lambda Q(z))}}{Z(\lambda)},
\label{ggibs}
\\
Z(\lambda)&=&\int_{\Gamma}  \e^{-\beta H(z,\lambda)}\,{\rm d}z, \qquad z \in \Gamma, \; \lambda \in \mathbb{R}.
\label{eq:partfunc}
\end{eqnarray} 
The logarithm of the partition function is related to the Helmholtz free energy $F$  by~\cite{landau}
\begin{equation}
\label{eq:Helmholtz}
F(\lambda)=-\beta^{-1}\ln Z(\lambda).
\end{equation}
Notice that for an unbounded phase space, such as $\Gamma= \mathbb{R}^{2d}$, the canonical state~\eqref{ggibs} is a well defined probability density if  $H(z,\lambda)$ is confining, that is, for all $\lambda\in\mathbb{R}$, $H(z,\lambda)\to + \infty$ (sufficiently fast) as $|z| \to \infty$.

Next, consider the time reversal transformation on the phase space $\Gamma$:
\begin{equation}
\theta :\Gamma \to \Gamma,\qquad  (q,p) \mapsto (q,-p).
\end{equation}
We assume that:
\begin{enumerate} 
\item the unperturbed Hamiltonian  $H_0$ is invariant under time-reversal transformations, say $H_0\left(\theta(z)\right)=H_0(q,-p)=H_0(z)$ for every point $z=(q,p)$ in phase space $\Gamma$;
\item the conjugate variable $Q$ has a definite behaviour under time-reversal transformations, say $Q(\theta(z))=\eta_Q Q(z)$, with  $\eta_Q=\pm 1$. (For example if $Q(q,p)=q_1$, then $\eta_Q=1$, while if $Q(q,p)=q_1 p_2-p_2q_1$, then $\eta_Q=-1$).
\end{enumerate}
As a consequence, we get that
\begin{equation}
\label{eq:timereversalcl}
H(\theta(z),\lambda)=  H(z,\eta_Q\lambda), 
\end{equation}
for all $z\in\Gamma$ and $\lambda\in\mathbb{R}$. Indeed,
\begin{equation}
H(\theta(z),\lambda)=H_0(\theta(z))-\lambda Q(\theta(z)) = H_0(z)-\lambda \eta_Q Q(z)  =  H(z,\eta_Q\lambda).
\end{equation}
In fact, condition~\eqref{eq:timereversalcl} is equivalent to the assumptions on $H_0$ and $Q$, as the reader can easily prove.

Suppose now that a given force \emph{protocol}, assumed to be smooth,
\begin{equation}
\label{eq:forceprotocol}
t \in [0,\tau] \mapsto \Lambda_t \in \mathbb{R}
\end{equation}
 is assigned to the external force $\lambda$ in~\eqref{eq:clham}, so that the Hamilton function $H(z,\Lambda_t)$ is time-dependent and the overall quantity $-\Lambda_t Q(z)$ is the time-dependent perturbation term.

Our intention is to implement a time-reversed protocol, since it is evidently impossible to turn back the physical time (a procedure that could be implemented on a computer simulation). Nevertheless, we need a way to implement a backward protocol in real time in order to deduce classical fluctuation relations. 
\begin{figure}[tbp]
\label{img:backwardprotocol}
\centering
\includegraphics[width=0.5\columnwidth]{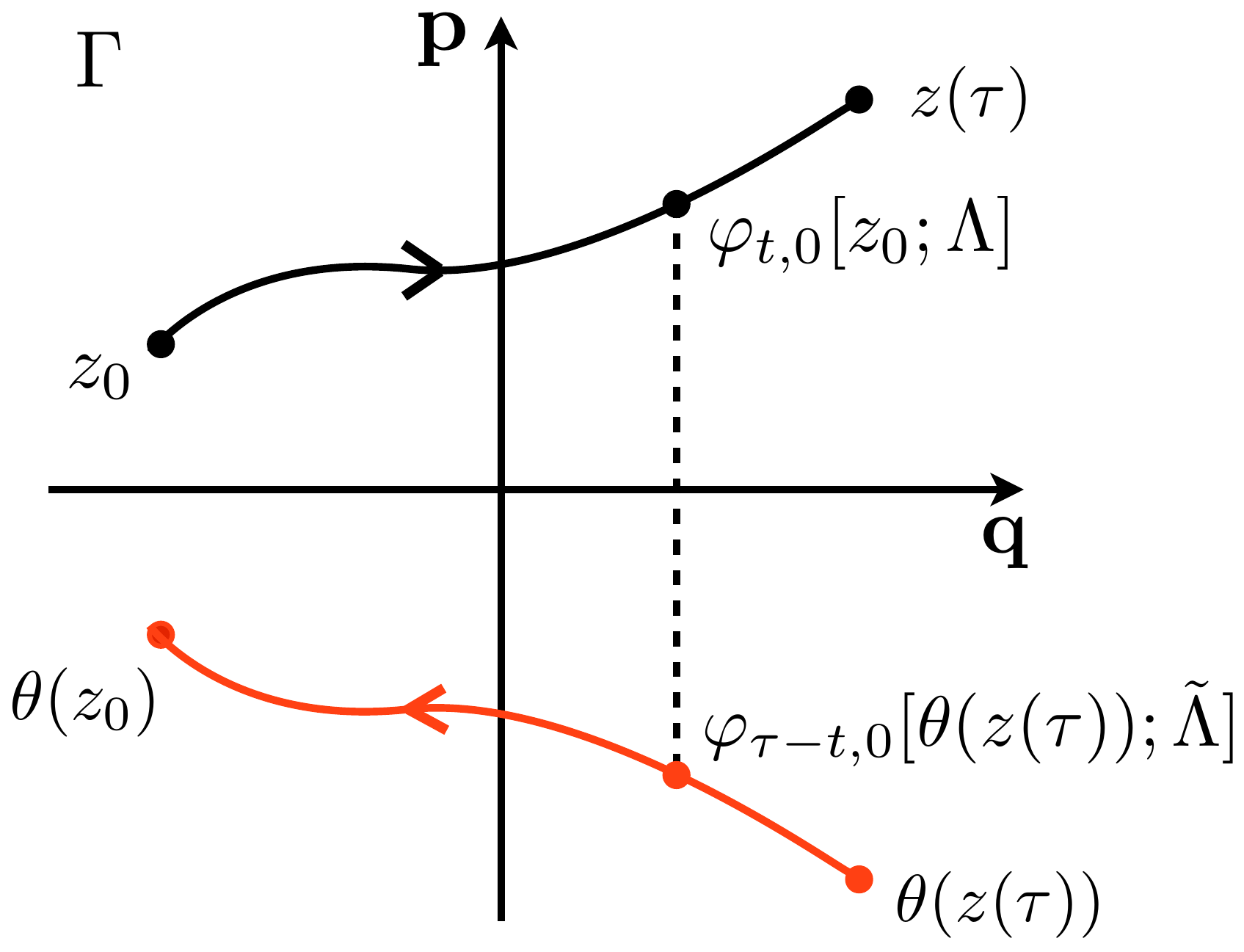}
\caption{ Microreversibility~{\protect\cite{campisi}. The initial point $z_0$ at time $t=0$ gets dragged under the protocol $\Lambda$ along the trajectory $\varphi_{t,0}[z_0;\Lambda]$, until it reaches its final position $z(\tau)$ at time $t=\tau$. Plotted in red is the trajectory under the backward  protocol $\tilde\Lambda$. It starts at $t=0$ in $\theta(z(\tau))$, evolves along $\varphi_{\tau-t,0}[\theta(z(\tau));\tilde\Lambda]$ until time $t=\tau$, at $\theta(z_0)$.}
\label{fig:timereversal}}
\end{figure}

We define the \emph{backward  protocol} as 
\begin{equation}
\label{eq:backward protocol}
\tilde\Lambda: [0,\tau] \to \mathbb{R}, \qquad \tilde\Lambda_t :=\eta_Q \Lambda_{\tau-t}.
\end{equation}
In this way, modulo a sign $\eta_Q$ related to the time-reversal parity of $Q$, the external force traces back its evolution from $\tilde\Lambda_0=\eta_Q \Lambda_{\tau}$ to $\tilde\Lambda_{\tau}=\eta_Q\Lambda_0$. 

We will denote by $\varphi_{t,0}[z_0;\Lambda]$ the solution, assumed to exist and to be unique, of the Hamilton equations at time $t$ under the external protocol $\Lambda$,
\begin{equation}
\begin{cases}
\dot q(t)=\nabla_p H(z(t),\Lambda_t),\\ \dot p(t)=-\nabla_q H(z(t),\Lambda_t),
\end{cases}
\label{eq:Hamiltoneqs}
\end{equation}
 with initial conditions $z(0)=z_0=(q_0,p_0)$, so that
\begin{equation}
z\in \Gamma \mapsto \varphi_{t,0}[z;\Lambda] \in \Gamma
\end{equation}
is the Hamiltonian flow in $[0,t]$ under the protocol $\Lambda$.

It is an instructive exercise on the use of Hamilton equations to prove that, under the microreversibility assumption~\eqref{eq:timereversalcl},
the flow generated by the backward protocol $\tilde\Lambda$ and the forward flow are related as follows~\cite{strato}:
\begin{equation}
\label{eq:fluxes}
\varphi_{t,0}[z_0;\Lambda]=\theta(\varphi_{\tau-t,0}[\theta(z(\tau));\tilde\Lambda]), \qquad z(\tau)= \varphi_{\tau,0}[z_0; \Lambda].
\end{equation}
Notice that $\theta(z(\tau))$ is the initial position under the backward protocol $\tilde\Lambda$. 
The lazy reader can convince himself of the validity of~\eqref{eq:fluxes} by looking at figure~\ref{fig:timereversal}.

The second ingredient is the initial state and its nature is statistical:
we assume that the initial conditions $z_0$ of the system are randomly sampled from the Gibbs canonical distribution~\eqref{ggibs} at $t=0$,
\begin{equation}
\rho_\beta^{\Lambda_0}(z)= \frac{ \e^{-\beta H(z, \Lambda_0)}}{Z(\Lambda_0)}.
\label{eq:incan}
\end{equation}
Then, we let the system evolve under the protocol $\Lambda$ until time $t=\tau$.

Since the dynamics is Hamiltonian, the \emph{work} $W[z_0;\Lambda]$ done by the external protocol on the system is given by the difference between the final energy and the initial one, namely
\begin{equation}
W[z_0;\Lambda]=H(z(\tau),\Lambda_\tau)-H(z_0,\Lambda_0),
\label{eq:workdefcl}
\end{equation}
where $z(\tau)= \varphi_{\tau,0}[z_0;\Lambda]$. Clearly, $W[z_0;\Lambda]$ depends on both the initial conditions $z_0$ and the protocol $\Lambda$.

The work can be written as
\begin{equation}
\label{eq:work}
W[z_0;\Lambda]=-\int_0^\tau\dot\Lambda_t \, Q(z(t))\,{\rm{d}}t,
\end{equation}
where $ \dot\Lambda_t= \frac{\rm d}{{\rm d}t} \Lambda_t$.
Indeed, we get
\begin{equation}
W[z_0;\Lambda]=H(z(\tau),\Lambda_\tau)-H(z_0,\Lambda_0)=\int_0^\tau \frac{\rm d}{{\rm d}t} H(z(t),\Lambda_t)\,{\rm d}t,
\end{equation}
where $z(t)= \varphi_{t,0}[z_0;\Lambda]$.
Next, we expand the total derivative of $H(z(t),\Lambda_t)$:
\begin{eqnarray}
\label{eq:timerev}
\frac{\rm d}{{\rm d}t} H(z(t),\Lambda_t)&=& \nabla_z H( z(t),\Lambda_t)\cdot \dot z(t) + \frac{\partial}{\partial\lambda} H(z(t),\Lambda_t)\,\dot \Lambda_t\\
&=& \frac{\partial}{\partial\lambda} H(z(t),\Lambda_t)\,\dot \Lambda_t = - Q(z(t))\,\dot \Lambda_t\,.
\end{eqnarray}
 In the second equality we used the orthogonality condition between the gradient of the Hamiltonian and the velocity vector field,
\begin{eqnarray}
\nabla_z H(z(t),\Lambda_t)\cdot \dot z(t) &=&\nabla_q H(z(t),\Lambda_t)\cdot\dot q (t)+\nabla_p H(z(t),\Lambda_t)\cdot\dot p(t)
\nonumber\\
&=& \nabla_q H (z(t),\Lambda_t)\cdot\nabla_p H (z(t),\Lambda_t)
\nonumber\\ 
& & -\nabla_p H(z(t),\Lambda_t)\cdot\nabla_q H(z(t),\Lambda_t) 
\nonumber\\
&=&0,
\end{eqnarray}
which is a direct consequence  of Hamilton equations~\eqref{eq:Hamiltoneqs}.
Thus, we have proved equation \eqref{eq:work}.
So far, we have only made use of Hamiltonian dynamics.

\subsection{Jarzynski equality}

We want to compute the average of $ \e^{-\beta W[z_0; \Lambda]}$ over all the possible initial conditions drawn from the canonical distribution~\eqref{eq:incan}:
\begin{equation}
\langle { \e^{-\beta W}}\rangle_{\Lambda}:=\int_\Gamma  \e^{-\beta W[z_0;\Lambda]}  \rho_\beta^{\Lambda_0}(z_0)\,{\rm d}z_0.
\label{eq:averexp}
\end{equation}
The Jarzynski equality states that 
\begin{equation}
\boxed{
\langle { \e^{-\beta W}}\rangle_{\Lambda}=  \e^{-\beta (F(\Lambda_{\tau})-F(\Lambda_0))}
}
\label{eq:Jarzeq}
\end{equation}
where $F$  is  the Helmoltz free energy~\eqref{eq:Helmholtz} 

This is a fluctuation relation since it connects a non-equilibrium quantity (on the left hand-side) with an equilibrium one (on the right hand side). The right hand side is an equilibrium quantity since it depends solely on the initial and final equilibrium states. Notice however that~\eqref{eq:Jarzeq} is exact: it holds no matter how strong is the external force $\Lambda$ and how far from the initial equilibrium $\rho_\beta^{\Lambda_0}$ the system is driven by $\Lambda$.

Let us consider the average~\eqref{eq:averexp}  and let us perform some simple computations:
\begin{eqnarray}
\label{eq:jar}
\langle { \e^{-\beta W}}\rangle_{\Lambda}&=&\int_\Gamma  \e^{-\beta W[z_0;\Lambda]} \rho_\beta^{\Lambda_0}(z_0)\,{\rm d}z_0
\nonumber\\
&=&\frac{1}{Z(\Lambda_0)}
\int_\Gamma  \e^{-\beta \left( H(z(\tau),\Lambda_\tau)- H(z_0,\Lambda_0)\right)}    \e^{-\beta H(z_0,\Lambda_0)} {\rm d}z_0,
\nonumber\\
&=&\frac{1}{Z(\Lambda_0)}
\int_\Gamma  \e^{-\beta H(z(\tau),\Lambda_\tau)}{\rm d}z_0.
\label{eq.35}
\end{eqnarray}

By the Liouville theorem, the Hamiltonian flow $\varphi_{\tau,0}[\,\cdot\,;\Lambda]$ is a canonical transformation on the phase space $\Gamma$, and thus is volume preserving and has unit  Jacobian determinant:
\begin{equation}
\label{eq:cantransf}
z=z(\tau)=\varphi_{\tau,0}[z_0;\Lambda],
\qquad 
\left|\frac{\partial z}{\partial z_0} \right| =1.
\end{equation}
Therefore the integration over the initial points $z_0$ in~\eqref{eq.35} can be traded for an integration over the final points $z$, namely
\begin{equation}
\frac{1}{Z(\Lambda_0)}\int_\Gamma  \e^{-\beta H(z(\tau),\Lambda_\tau)}{\rm d}z_0=
\frac{1}{Z(\Lambda_0)}\int_\Gamma  \e^{-\beta H(z,\Lambda_\tau)}{\rm d}z\overset{\eqref{eq:partfunc}}{=}
\frac{Z(\Lambda_\tau)}{Z(\Lambda_0)}.
\end{equation}
Thus we get
\begin{equation}
\langle { \e^{-\beta W}}\rangle_{\Lambda}=\frac{Z(\Lambda_\tau)}{Z(\Lambda_0)},
\end{equation}
which gives the Jarzynski equality by using the definition of the Helmholtz free energy~\eqref{eq:Helmholtz}.

\begin{figure}[tbp]
\centering
\includegraphics[width=0.5\columnwidth]{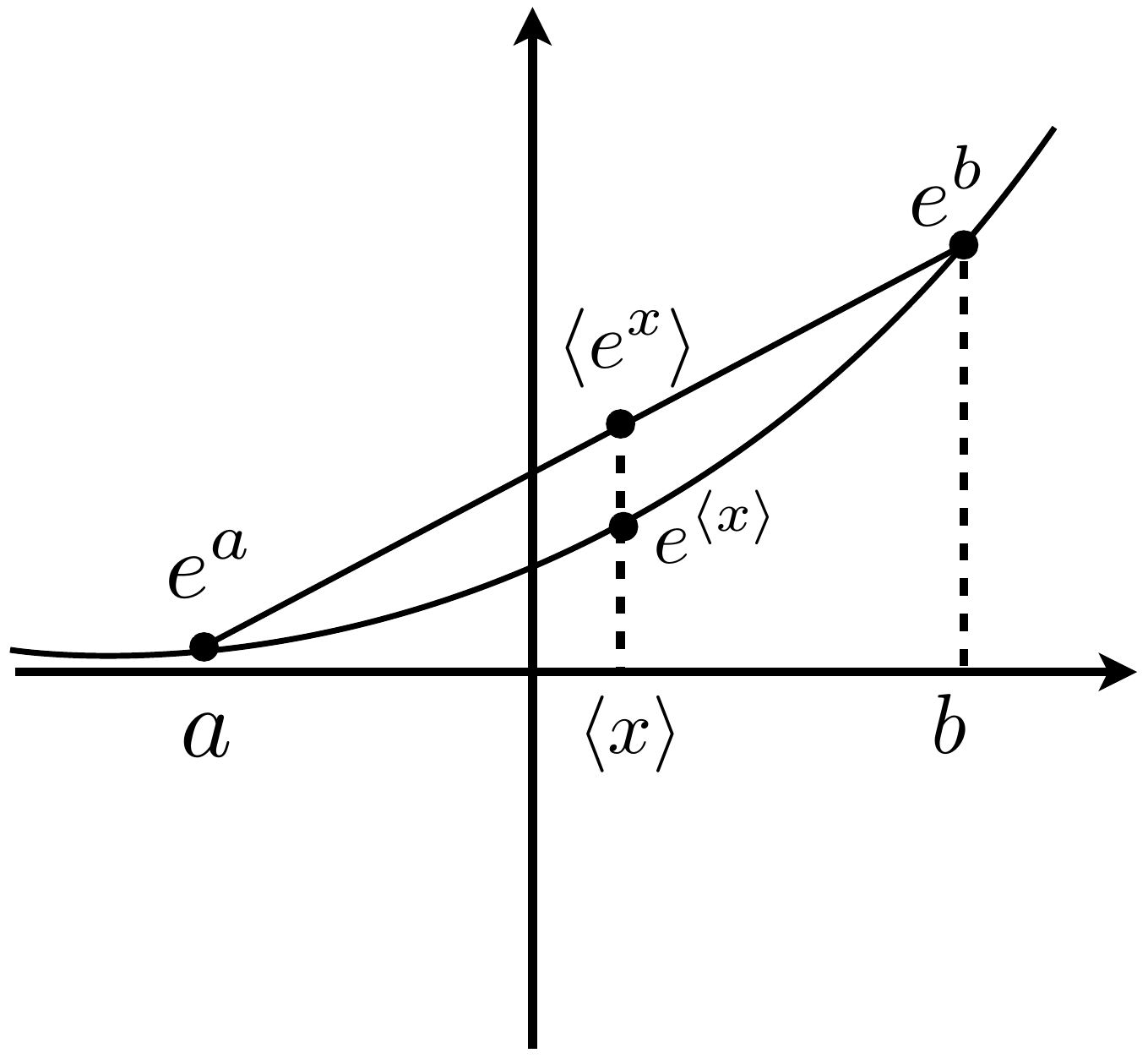}
\caption{Given two points $a,b$ and denoting the average with $\langle x\rangle=(a+b)/2$, from the convexity of the exponential function it follows that $\langle e^x\rangle\ge  \e^{\langle x\rangle}$, where  $\langle e^x\rangle=(e^a+e^b)/2$ is  the average of the exponential function over the points $a$ and $b$. \label{img:convexexp} }
\end{figure}

A straightforward consequence of~\eqref{eq:Jarzeq} follows from the convexity of the exponential function, see Figure \ref{img:convexexp}. By Jensen's inequality we get
\begin{equation}
 \e^{-\beta(F(\Lambda_\tau)-F(\Lambda_0))}=\langle { \e^{-\beta W}}\rangle_{\Lambda} \ge  \e^{-\beta \langle W\rangle_{\Lambda}}.
\end{equation}
Therefore, one has
\begin{equation}\label{eqn:Jarzynskiineq}
\langle W\rangle_\Lambda \ge  F(\Lambda_\tau)-F(\Lambda_0)=:\Delta F, 
\end{equation}
which is an expression of the second law of thermodynamics.
Indeed, if one defines the dissipated work as~\cite{landau}
\begin{equation}
W_{\rm diss}=W-\Delta F,
\end{equation} 
then  inequality~\eqref{eqn:Jarzynskiineq} states that the dissipated work on average can only be absorbed: 
\begin{equation}
\langle W_{\rm diss}\rangle_{\Lambda}=\langle W\rangle_{\Lambda}-\Delta F \ge0.
\end{equation}
In other words the  dissipated work done on a system initially at thermal equilibrium is always nonnegative, independently of the protocol $\Lambda$.

\subsection{Crooks fluctuation theorem}

Next, we will consider the Crooks fluctuation theorem and deduce from it the  Jarzynski equality. In order to do so, we need to introduce probability density functions and make use of the microreversibility assumption~\eqref{eq:timereversalcl}.

The probability density function (from now on PdF) of the work under the protocol $\Lambda$ is given by
\begin{equation}
p[ w;\Lambda]=\langle\delta( w-W)\rangle_{\Lambda},
\end{equation}
where the average is over the initial Gibbs ensemble $\rho_\beta^{\Lambda_0}$ in~\eqref{eq:incan}, namely 
\begin{equation}
p[ w;\Lambda]=\int_{\Gamma} \frac{ \e^{-\beta H(z_0, \Lambda_0)}}{Z(\Lambda_0)} \delta( w-W[z_0;\Lambda]) \, {\rm d}z_0,
\label{eq:pwLambdadef}
\end{equation}
and $\delta$ is the Dirac measure.
From the PdF of the work one can get the average of any continuous bounded function $f$. Indeed,
\begin{eqnarray}
\langle  f(W) \rangle_{\Lambda} &=& \int_{\Gamma}   f(W[z_0;\Lambda ]) \frac{ \e^{-\beta H(z_0,\Lambda_0)}}{Z(\Lambda_0)} \,{\rm d} z_0
\nonumber \\
&=&\int_{\Gamma} {\rm d} z_0 \, \frac{ \e^{-\beta H(z_0,\Lambda_0)}}{Z(\Lambda_0)} \int_{\mathbb{R}}{\rm d}  w \, f(w) \delta( w-W[z_0;\Lambda ]) 
\nonumber \\
&=&\int_{\mathbb{R}}{\rm d}  w \, f(w) \int_{\Gamma} {\rm d} z_0 \,  \frac{\e^{-\beta H(z_0,\Lambda_0)}}{Z(\Lambda_0)}\delta( w-W[z_0;\Lambda ]) \nonumber \\
&=&\int_{\mathbb{R}} f(w) p[ w;\Lambda] \,{\rm d}  w. 
\label{eqn:expval}
\end{eqnarray}

Our objective now is to prove the Crooks fluctuation theorem:
\begin{equation}\label{eq:Crooks}
\boxed{
p[ w;\Lambda]= \e^{\beta ( w-\Delta F)}\,p[- w; \tilde\Lambda],}
\end{equation}
where
\begin{equation}
\Delta F=F(\Lambda_\tau)-F(\Lambda_0)=-\frac{1}{\beta}\ln \left( \frac{Z(\Lambda_\tau)}{Z(\Lambda_0)}\right).
\end{equation}
The above equality relates the probability of absorbing work under the protocol $\Lambda$ to that of releasing work under the backward protocol $\tilde\Lambda$. It is worth noticing  that the ratio between the two probabilities is exponentially small.

We start with a simple manipulation of the PdF~\eqref{eq:pwLambdadef}:
\begin{eqnarray}
p[ w;\Lambda]&=&\int_\Gamma \frac{\e^{-\beta H(z_0,\Lambda_0)}}{Z(\Lambda_0)} \,\delta( w-W[z_0;\Lambda])\,{\rm d}z_0 \nonumber \\
&=&\frac{1}{Z(\Lambda_0)}\e^{\beta  w}\int_\Gamma   \e^{-\beta H(z(\tau),\Lambda_\tau)} \delta( w-H(z(\tau),\Lambda_\tau)+H(z_0,\Lambda_0))\,{\rm d}z_0 \nonumber \\
&=&\frac{Z(\Lambda_\tau)}{Z(\Lambda_0)}\e^{\beta  w}\int_\Gamma  \frac{\e^{-\beta H(z(\tau),\Lambda_\tau)}}{Z(\Lambda_\tau)}  \delta(- w-H(z_0,\Lambda_0)+H(z(\tau),\Lambda_\tau))\,{\rm d}z_0 \nonumber \\
&=&\e^{\beta(  w-\Delta F)}\int_\Gamma \rho_\beta^{\Lambda_\tau} (z(\tau))\delta(- w-H(z_0,\Lambda_0)+H(z(\tau),\Lambda_\tau))\,{\rm d}z_0 ,
\label{eq:345*}
\end{eqnarray}
where, we used the definition of work~\eqref{eq:workdefcl} and, in the last equality, the definition of $\rho_\beta^{\Lambda_\tau} (z)$ in equation~\eqref{ggibs}. 

Next we add the ingredient of microreversibility in order to rewrite the above integral. 
We recall the relation between the Hamiltonian flow under the protocol $\Lambda$ and the one generated by the backward protocol $\tilde\Lambda$ given in equation~\eqref{eq:fluxes}, which is valid for all $t \in [0, \tau]$.
Thus, at time $t=0$ we get an alternative way of writing the initial datum $z_0$:
\begin{equation}
\label{eq:flux1}
z_0=\varphi_{0,0}[z_0;\Lambda]= \theta(\varphi_{\tau,0}[\theta(z(\tau)); \tilde\Lambda]).
\end{equation}
Using the microreversibility assumption~\eqref{eq:timereversalcl}, 
we can rewrite the Hamiltonian at $t=0$ as follows:
\begin{eqnarray}
H(z_0,\Lambda_0)&=&H(\theta(\varphi_{\tau,0}[\theta(z(\tau));  \tilde\Lambda]),\Lambda_0) 
=H(\varphi_{\tau,0}[\theta(z(\tau));  \tilde\Lambda],\eta_Q\Lambda_0) 
\nonumber\\
&=&H(\varphi_{\tau,0}[\theta(z(\tau));  \tilde\Lambda], \tilde\Lambda_\tau), 
\end{eqnarray}
since $\tilde\Lambda_\tau=\eta_Q \Lambda_0$.
Moreover the same assumption implies that  
\begin{equation}
\label{eq:fintime}
H(z(\tau),\Lambda_\tau)=H(\theta(z(\tau)), \tilde \Lambda_0),
\end{equation}
because $\tilde \Lambda_0=\eta_Q \Lambda_\tau$.
We are now ready to plug equation \eqref{eq:flux1}-\eqref{eq:fintime} into  equation \eqref{eq:345*} 
and the conservation of measure induced by the transformation in equation~\eqref{eq:cantransf},
obtaining
\begin{eqnarray}
\frac{ p[ w;\Lambda] }{ \e^{\beta(  w-\Delta F)}}
&=&\int_\Gamma \rho_\beta^{\Lambda_\tau} (z(\tau))\delta(- w-H(z_0,\Lambda_0)+H(z(\tau),\Lambda_\tau))\,{\rm d}z_0 \nonumber \\
&=& \int_\Gamma \rho_\beta^{\Lambda_\tau} (z(\tau))\delta(- w-H(\varphi_{\tau,0}[\theta(z(\tau));\tilde\Lambda],\, \tilde\Lambda_\tau)+H(\theta(z(\tau)) , \tilde \Lambda_0))\,{\rm d}z_0\nonumber \\
&=&  \int_\Gamma \rho_\beta^{\Lambda_\tau} (\theta(z))\delta(- w-H(\varphi_{\tau,0}[z;\tilde\Lambda],\, \tilde\Lambda_\tau)+H(z, \tilde \Lambda_0))\,{\rm d}z \label{eq:changez}
\nonumber\\
&=&  \int_\Gamma \rho_\beta^{ \tilde\Lambda_0} (z)\delta(- w-H(\varphi_{\tau,0}[z;\tilde\Lambda],\, \tilde\Lambda_\tau)+H(z, \tilde \Lambda_0))\,{\rm d}z \label{eq:timerevinv}
\nonumber\\
&=& p[- w; \tilde\Lambda] 
\end{eqnarray}
where in \eqref{eq:changez} we performed the change of coordinate induced by the time reversal transformation $ z=\theta(z(\tau))$ (see equation \eqref{eq:timerev}), whose Jacobian has unit determinant, and we used the relation
\begin{equation}
\rho_\beta^{\Lambda_\tau} (\theta(z))=\frac{\e^{-\beta H(\theta(z), \Lambda_\tau)}}{Z(\Lambda_\tau)}=\frac{\e^{-\beta H(z, \eta_Q \Lambda_\tau)}}{Z(\eta_Q \Lambda_\tau)}=\rho_\beta^{\eta_Q \Lambda_\tau}(z)=\rho_\beta^{ \tilde\Lambda_0}(z).
\end{equation}

The Crooks fluctuation relation~\eqref{eq:Crooks} states that, if we consider a positive work $ w>\Delta F>0$, the probability that the work is injected into the system is larger by a factor $\e^{\beta( w-\Delta F)}$ than the probability that it might be absorbed under the backward forcing. In other words, energy consuming processes are exponentially more probable than energy releasing processes. Crooks fluctuation theorem expresses the second law of thermodynamics at a detailed level which quantifies the relative frequency of energy releasing processes.

Moreover, from the Crooks fluctuation theorem it is possible to recover the Jarzynski equality, in fact from \eqref{eqn:expval} and \eqref{eq:Crooks} one has
\begin{eqnarray}
\langle  \e^{-\beta W} \rangle_{\Lambda}  
&=&\int_{\mathbb{R}} \e^{-\beta   w} p[ w;\Lambda] {\rm d}  w 
=\int_{\mathbb{R}} \e^{-\beta   w} \e^{\beta(  w-\Delta F)} p[- w; \tilde\Lambda] {\rm d}  w 
\nonumber\\
&=& \e^{-\beta \Delta F}\int_{\mathbb{R}}  p[- w; \tilde\Lambda] {\rm d}  w 
= \e^{-\beta \Delta F} ,
\end{eqnarray}
since $\int_{\mathbb{R}} p[- w; \tilde\Lambda] {\rm d}  w=1$.

A final comment is in order. A straightforward corollary of the Crooks fluctuation relation~\eqref{eq:Crooks} is the following generalization of the Jarzinski equality, whose proof is left to the reader.

Let $B: \Gamma \to \mathbb{R}$ such that  $B(\theta(z))=\eta_B B(z)$, for all $z \in \Gamma$, with $\eta_B=\pm 1$, then
\begin{equation}
\left\langle  \exp\left( \int_0^\tau u_t B(t) \, {\rm d}t\right)   \e^{-\beta  W }\right\rangle_{\Lambda}= \e^{-\beta \Delta F} \left\langle  \exp\left( \int_0^\tau \tilde u_t  B(t) \, {\rm d}t\right) \right\rangle_{ \tilde \Lambda},
\label{eq:flucdissallord}
\end{equation}
for all test functions $u:[0,\tau] \to \mathbb{R}$, where $\tilde u_t:=\eta_B u_{\tau-t}$, for all $t \in [0,\tau]$, is the backward function. The relation~\eqref{eq:flucdissallord} is the generating functional of the fluctuation-dissipation relations at all order: take the functional derivatives with
respect to $\Lambda$ and $u$ at $\Lambda=u\equiv0$.

\section{Lecture 3: Quantum Fluctuation Relations}
In this lecture we would like to discuss the quantum version of the fluctuation relations, whose classical version were proved in the previous section.

From the axioms of quantum mechanics \cite{von} the Hamiltonian function on phase space has to be substituted with a self-adjoint operator. For this reason we are going to consider the following Hamiltonian operator on the Hilbert space $\mathcal{H}$:
\begin{equation}
\label{eq:ham}
H(\lambda)= H_0-\lambda Q,
\end{equation}
where $H_0$ and $Q$ are self-adjoint operators, while $\lambda$ is a real parameter representing the external force. 
The canonical Gibbs state is described quantum mechanically by a density matrix \cite{huang}:
\begin{equation}
\rho_\beta^\lambda=\frac{1}{Z(\lambda)}\e^{-\beta H(\lambda)},
\end{equation}
where 
\begin{equation}
Z(\lambda)=\tr(\e^{-\beta H(\lambda)})
\end{equation} is the partition function. The Helmholtz free energy is defined in terms of the partition function as in the classical case~\eqref{eq:Helmholtz}.

In the following we will assume that $\{H(\lambda)\}_{\lambda\in\mathbb{R}}$ is a family of (unbounded) self-adjoint operators on a common domain $D$.
Moreover, we assume that $\rho_\beta^\lambda$ is trace-class for all  $\lambda\in\mathbb{R}$ and $\beta>0$.
This implies that the Hamiltonians~\eqref{eq:ham} have a discrete spectrum with finite multiplicity,  namely,
\begin{equation}
H(\lambda)=\sum_{m}E_m^\lambda\Pi_m^\lambda,
\label{eq:specdec}
\end{equation}
where $\{E_m^\lambda\}$ are the distinct eigenvalues of $H(\lambda)$ ($E_m^\lambda\neq E_n^\lambda$ for $m\neq n$),  and the eigenprojections $\{\Pi_m^\lambda\}$ are of finite rank, $\tr (\Pi_m^\lambda)<+\infty$ for all $m$. Moreover, if $\mathcal{H}$ is infinite-dimensional, then $E_m^\lambda \to +\infty$ as $m\to +\infty$. 

Assume now that a given protocol  $t \in [0,\tau] \mapsto \Lambda_t \in \mathbb{R}$ is assigned to the external force in~\eqref{eq:ham}, so that the Hamiltonian is time-dependent, $t\mapsto H(\Lambda_t)$. The quantum evolution in the interval $[s,t]$ is described by the Schr\"odinger equation, which, from the operator point of view, reads
\begin{equation}
\ii \frac{\partial}{\partial t} U_{t,s}[\Lambda] \psi =H(\Lambda_t) U_{t,s}[\Lambda] \psi, \qquad U_{s,s}[\Lambda]=\mathbb{I}, \qquad (t\geq s),
\label{eq:UF}
\end{equation}
for all $\psi\in D$:
if the system is initially in the state $\psi \in D$, according to the Schr\"odinger equation it will evolve at time $t$ to the state $U_{t,s}[\Lambda]\psi \in D$.
Notice that one can instead consider the derivative with respect to the initial time $s$ and get
\begin{equation}
\ii \frac{\partial}{\partial s} U_{t,s}[\Lambda] \psi= U_{t,s}[\Lambda] H(\Lambda_{s}) \psi, \qquad U_{t,t}[\Lambda]=\mathbb{I}, \qquad (t\geq s),
\label{eq:UB}
\end{equation}
for all $\psi\in D$, which is a final value problem. 
We will explicitly denote the unique solution to~\eqref{eq:UF} or to~\eqref{eq:UB} by
\begin{equation}
U_{t,s}[\Lambda]= \mathcal{T}\exp \bigl(-\ii \int_s^t H(\Lambda_u)\d u\bigr)=\mathcal{T} \exp \bigl(-\ii \int_s^t \left(H_0-\Lambda_u Q \right) \d u\bigr),
\end{equation}
where $\mathcal{T}$ is the time-ordered product.

In analogy with the classical case~\eqref{eq:workdefcl}, let us tentatively define the work as the difference between the Hamiltonian operator at time $t=\tau$ and the Hamiltonian operator at time $t=0$ in the Heisenberg picture:
\begin{equation}
\label{eqwork}
W[\Lambda]=U^\dagger_{\tau,0}[\Lambda] H(\Lambda_\tau)U_{\tau,0}[\Lambda]-H(\Lambda_0).
\end{equation}
In order to get Jarzynski equality, then, one could try to follow step by step the derivation used for the classical case (see equation \eqref{eq:jar}), but in this case the cancellation in equation \eqref{eq:jar} cannot be made. Indeed, in general the Hamiltonians~\eqref{eq:ham} at different times do not commute, unless $H_0$ does not commute with $Q$:
\begin{equation}
[H(\Lambda_t),H(\Lambda_s)]= (\Lambda_t-\Lambda_s)[H_0,Q].
\end{equation}
 In fact, it can be shown that, with the definition of work given by~\eqref{eqwork}, one has that
\begin{equation}
\langle {\e^{-\beta W}}\rangle_{\Lambda}:= {\rm tr} (\rho_\beta^{\Lambda_0} \e^{- \beta W[\Lambda]})=\e^{-\beta \Delta F}
\end{equation}
if and only if $[H(\Lambda_t),H(\Lambda_s)]=0$ for all $t, s \in [0,\tau]$. The last condition applies either to the case of a constant protocol, which would  imply $\Delta F=0$ or to the commutative case $[H_0,Q]=0$ that is, morally, to a classical situation.

At first look it may seem that there could be no quantum counterpart of the Jarzynski equality. 
The problem, in fact,  relies on the definition~\eqref{eqwork} of work under the protocol $\Lambda$, and one has to think more deeply about the meaning of work.

It is well known that work characterizes a process rather than a state of the system, and indeed it depends on the trajectory followed by the system from its initial to its final state (see equation~\eqref{eq:workdefcl}). As such, in quantum mechanics, work cannnot be associated to a self-adjoint operator whose eigenvalues are determined by an  energy measurement at a given time. Instead, in order to determine the exchanged work, energy must be measured twice, at two instants of time, say at time $t=0$ and time $t=\tau$. The difference of the outcomes of these two measurements will yield the work performed on the system in that particular instance.

This two-time measurements procedure is the operational description of the measure of work~\eqref{eq:workdefcl} along a single trajectory $z_0 \mapsto z(\tau)$, which in the classical case is a particular instance of all possible trajectories drawn from the initial thermal state~\eqref{eq:incan}.
This description can be immediately exported to the quantum world, where, however, the measurement process will add quantum fluctuations to the classical statistical fluctuations due to the choice of a thermal initial state. As a consequence, the difference of the outcomes of a two-time measurement is different from the measurement of the difference between the corresponding measured operators~\eqref{eqwork}, whose operational meaning is unclear.

Let us look step by step at the two-measurement process:
\begin{enumerate}

\item 
As in the classical case the system is prepared in the  Gibbs state
\begin{equation}
\label{init}
\rho_\beta^{\Lambda_0}=\frac{\e^{-\beta H(\Lambda_0)}}{Z(\Lambda_0)}=\frac{1}{Z(\Lambda_0)}\sum_{n}\e^{-\beta E_n^{\Lambda_0}}\Pi_n^{\Lambda_0}.
\end{equation}

\item
At time $t=0$ the energy of the system is measured and the outcome is, say, the eigenvalue $E_n^{\Lambda_0}$ for some $n$, and 
the state of the system becomes 
\begin{equation}
\rho_n=\frac{\Pi^{\Lambda_0}_n\rho_\beta^{\Lambda_0}\Pi^{\Lambda_0}_n}{p_n^{\Lambda_0}},\qquad p_n^{\Lambda_0}=\tr (\rho_\beta^{\Lambda_0}\Pi^{\Lambda_0}_n )
\end{equation}
where $p_n^{\Lambda_0}=\tr (\rho_\beta^{\Lambda_0}\Pi^{\Lambda_0}_n )$ is the probability of getting the outcome~$E_n^{\Lambda_0}$.

\item 
Then,  one lets the system evolve for a time $\tau$, so that its state becomes
\begin{equation}
\rho_n(\tau)=U_{\tau,0}[\Lambda] \rho_n U^\dagger_{\tau,0}[\Lambda].
\end{equation}

\item 
Finally, at time $t=\tau$ a second energy  measurement is performed and the outcome  $E_m^{\Lambda_\tau}$ is obtained with probability 
\begin{equation}
p_{m|n}=\tr  (\Pi_m^{\Lambda_\tau}\rho_n(\tau)),
\end{equation}
and the state of the system becomes
\begin{equation}
\rho_{m,n}=\frac{\Pi^{\Lambda_\tau}_m U_{\tau,0}[\Lambda]\rho_n U^\dagger_{\tau,0}[\Lambda] \Pi^{\Lambda_\tau}_m}{p_{m|n}} 
=\frac{\Pi^{\Lambda_\tau}_m U_{\tau,0}[\Lambda]\Pi^{\Lambda_0}_n\rho_\beta^{\Lambda_0}\Pi^{\Lambda_0}_n U^\dagger_{\tau,0}[\Lambda] \Pi^{\Lambda_\tau}_m}{p_{m|n}p_n}.
\label{eq:rhomn}
\end{equation}

\end{enumerate}

Summarizing, the overall work done on the system in this particular instance is given by the difference of the final and initial measurement outcomes,
\begin{equation}
W=E_m^{\Lambda_\tau}-E_n^{\Lambda_0},
\label{eq:eigendiff}
\end{equation} 
whose probability is  
\begin{eqnarray}
p_{m,n} &=& p_{m|n}  p_n^{\Lambda_0}\nonumber\\
&=&\tr \left(\Pi^{\Lambda_\tau}_m\rho_n(\tau)\right) \tr \left(\rho_\beta^{\Lambda_0}\Pi^{\Lambda_0}_n\right)\nonumber\\
&=& \tr \left(\Pi^{\Lambda_\tau}_m U_{\tau,0}[\Lambda]\frac{\Pi^{\Lambda_0}_n\rho_\beta^{\Lambda_0}\Pi^{\Lambda_0}_n}{p_n^{\Lambda_0}}U^\dagger_{\tau,0}[\Lambda]\right)p_n^{\Lambda_0}\nonumber\\
\label{eq:trace}
&=& \tr  \left(U^\dagger_{\tau,0}[\Lambda]\Pi^{\Lambda_\tau}_mU_{\tau,0}[\Lambda]\Pi^{\Lambda_0}_n\rho_\beta^{\Lambda_0}\Pi^{\Lambda_0}_n\right),
\end{eqnarray}
where, in the last equality, the cyclicity of the trace was used.
From equation~\eqref{init} it follows that
\begin{equation}
\Pi_n^{\Lambda_0}\rho_\beta^{\Lambda_0}\Pi_n^{\Lambda_0}=\frac{1}{Z(\Lambda_0)}\e^{-\beta E_n^{\Lambda_0}}\Pi_n^{\Lambda_0},
\end{equation}
so that the probability of obtaining the outcomes $(E_n^{\Lambda_0}, E_m^{\Lambda_\tau})$, and thus the state~\eqref{eq:rhomn}, reads
\begin{equation}
p_{m,n}= \frac{\e^{-\beta E_n^{\Lambda_0}}}{Z(\Lambda_0)}\tr  \left(U^\dagger_{\tau,0}[\Lambda]\Pi^{\Lambda_\tau}_mU_{\tau,0}[\Lambda]\Pi^{\Lambda_0}_n \right).
\label{eq:pmn}
\end{equation}

Now we can compute the average of $\e^{-\beta W}$ over all possible outcomes:
\begin{eqnarray}
\langle \e^{-\beta W}\rangle_{\Lambda} & := & \sum_{m,n}\e^{-\beta (E_m^{\Lambda_\tau}-E_n^{\Lambda_0})}
p_{m,n}\nonumber \\
& = & \sum_{m,n} \frac{\e^{-\beta E_m^{\Lambda_\tau}}}{Z(\Lambda_0)}\tr  \left(U^\dagger_{\tau,0}[\Lambda]\Pi^{\Lambda_\tau}_mU_{\tau,0}[\Lambda]\Pi^{\Lambda_0}_n \right) \nonumber \\
&=&\sum_m \frac{\e^{-\beta E_m^{\Lambda_\tau}}}{Z(\Lambda_0)}\tr  \left(U^\dagger_{\tau,0}[\Lambda]\Pi^{\Lambda_\tau}_mU_{\tau,0}[\Lambda]\sum_n\Pi^{\Lambda_0}_n \right)  \label{eqn:id+spectral} \nonumber \\
&=&\sum_m\frac{\e^{-\beta E_m^{\Lambda_\tau}}}{Z(\Lambda_0)}\tr  \left(\Pi^{\Lambda_\tau}_m\right) = \frac{1}{Z(\Lambda_0)}\tr \left(\sum_m \e^{-\beta E_m^{\Lambda_\tau}}\Pi^{\Lambda_\tau}_m\right) \label{eqn:id+cycl} \nonumber\\
&=& \frac{1}{Z(\Lambda_0)}\tr \left(\e^{-\beta H(\Lambda_\tau)}\right) = \frac{Z(\Lambda_\tau)}{Z(\Lambda_0)}
= \e^{-\beta \Delta F} 
\end{eqnarray}
where, 
we used the cyclicity of the trace, the relation
\begin{equation}
\sum_m\tr (\Pi^{\Lambda_0}_m A)= \tr \left(\sum_m \Pi^{\Lambda_0}_m A\right) = \tr (A),
\end{equation} 
valid for all trace-class operators $A$, and  the equality 
\begin{equation}
\e^{-\beta H(\Lambda_\tau)}=\sum_{m}\e^{-\beta E_m^{\Lambda_\tau}}\Pi_m^{\Lambda_\tau}.
\end{equation}
Thus we have proved the Jarzynski equality~\eqref{eq:Jarzeq} for a quantum system.

This equality takes into account the presence of thermal fluctuations of work due to the Gibbs initial state, as in the classical case, together with
its quantum fluctuations inherent in the two-time measurement process.

\subsection{Microreversibility}
Let $\Theta: \mathcal{H} \to \mathcal{H}$ be the quantum time-reversal (anti-unitary) operator, with $\Theta^2= \mathbb{I}$. We assume that the Hamiltonian is time-reversal invariant, namely that for all $\lambda\in\mathbb{R}$:
\begin{equation}
\Theta H(\lambda) \Theta = H(\eta_Q \lambda),
\label{eq:timerevinv1}
\end{equation}
which is the quantum version of the assumption~\eqref{eq:timereversalcl}. As in the classical case, $\eta_Q$ is the time-reversal parity of the observable $Q$, namely 
\begin{equation}
\Theta Q \Theta = \eta_Q Q, \qquad \eta_Q=\pm1.
\end{equation}
From the spectral decomposition~\eqref{eq:specdec}, one gets that the property~\eqref{eq:timerevinv1} implies that
\begin{equation}
\label{eq:timereveigen}
E_m^{\eta_Q\lambda} = E_m^\lambda, \qquad \Theta \Pi_m^{\lambda} \theta = \Pi_m^{\eta_Q \lambda},
\end{equation}
for all $m$ and $\lambda$.

We will prove that
\begin{equation}\label{eqn:qmirev}
U_{t,0}[\Lambda]= \Theta U_{\tau-t}[\tilde{\Lambda}]\Theta U_{\tau,0}[\Lambda],
\end{equation}
where $t \in [0,\tau] \mapsto \tilde{\Lambda}_t:= \eta_Q \Lambda_{\tau-t}$ is the backward protocol~\eqref{eq:backward protocol}.
Notice the analogy with the classical case where
\begin{equation}
\varphi_{t,0}[\,\cdot\,;\Lambda]=\theta(\varphi_{\tau-t,0}[\theta(\varphi_{\tau,0}[\,\cdot\,;\Lambda]);  \tilde{\Lambda}]),
\end{equation}
see equation \eqref{eq:fluxes}. In order to prove \eqref{eqn:qmirev} we observe that, from~\eqref{eq:UF}, $U_{\tau-t,0}[\tilde{\Lambda}]$ satisfies the following integral equation on the common domain $D$:
\begin{eqnarray}
U_{\tau-t,0}[\tilde{\Lambda}]&=& \mathbb{I} - \ii \int_0^{\tau-t} \mathrm{d}s\, H(\tilde{\Lambda}_s)  U_{s,0}[\tilde{\Lambda}]
\nonumber\\
                                        &=&\mathbb{I} - \ii \int_0^{\tau-t} \mathrm{d}s\, H(\eta_Q\Lambda_{\tau-s})  U_{s,0}[\tilde{\Lambda}] \nonumber \\
                                        &=& \mathbb{I} - \ii \int_t^{\tau} \mathrm{d}\sigma\, H(\eta_Q \Lambda_{\sigma})  U_{\tau - \sigma,0}[\tilde{\Lambda}] .
\end{eqnarray}
Therefore, by using~\eqref{eq:timerevinv1}, one has
\begin{eqnarray}
\Theta U_{\tau-t,0}[\eta_Q\tilde{\Lambda}] \Theta 
&=& \mathbb{I} +\ii \int_t^{\tau} \mathrm{d}\sigma\, \Theta H( \eta_Q \Lambda_{\sigma})  U_{\tau - \sigma,0}[\tilde{\Lambda}] \Theta 
\nonumber\\
&=& \mathbb{I} + \ii \int_t^{\tau} \mathrm{d}\sigma\,  H( \Lambda_{\sigma})  \Theta U_{\tau - \sigma,0}[\tilde{\Lambda}] \Theta .
\label{eq:TUT}
\end{eqnarray}
On the other hand, from~\eqref{eq:UB}, one gets
\begin{equation}
U_{\tau, t} [\Lambda] = \mathbb{I} - \ii \int_t^\tau \mathrm{d}\sigma\, U_{\tau,\sigma}[\Lambda] H(\Lambda_\sigma)
\end{equation}
and thus
\begin{equation}
U_{\tau, t}^\dagger [\Lambda] = \mathbb{I} + \ii \int_t^\tau \mathrm{d}\sigma\, H( \Lambda_\sigma) U_{\tau,\sigma}^\dagger[ \Lambda] .
\label{eq:Udag}
\end{equation}
By comparing~\eqref{eq:TUT} with~\eqref{eq:Udag} we see that the unitaries $\Theta U_{\tau-t,0}[\tilde{\Lambda}] \Theta$ and  $U_{\tau, t}^\dagger[\Lambda]$ satisfy the same integral equation, whence by uniqueness they are equal:
\begin{equation}
\Theta U_{\tau-t,0}[\tilde{\Lambda}] \Theta = U_{\tau, t}^\dagger [\Lambda]= U_{t,0}[\Lambda]U_{\tau,0}^\dagger[\Lambda].
\end{equation}

\subsection{Quantum Crooks fluctuation theorem}
The probability $p_{m,n}$ of getting the outcomes $(E_n^{\Lambda_0}, E_m^{\Lambda_\tau})$ in the two-measurement process of the work done on a quantum system was derived in~\eqref{eq:pmn}. 
It follows that  the probability of getting a value $w$ of the  work is 
\begin{equation}
P[ w; \Lambda]= \sum_{m,n}p_{m,n} \delta( w, E^{\Lambda_\tau}_m - E^{\Lambda_0}_n ),
\end{equation}
where $\delta(x,y)$ is the Kronecker delta.
We want to prove that
\begin{equation}\label{eq:QCrooks}
\boxed{
P[ w;\Lambda]= \e^{\beta ( w-\Delta F)} P[- w; \tilde\Lambda].}
\end{equation}
We first observe that
\begin{eqnarray}
P[ w;\Lambda] &=& \sum_{m,n}p_{m,n} \delta( w, E^{\Lambda_\tau}_m - E^{\Lambda_0}_n) \nonumber \\
                                      &=& \sum_{m,n}    \frac{\e^{-\beta E^{\Lambda_0}_n}}{Z(\Lambda_0)}  \tr  \left(U^\dagger_{\tau,0}[\Lambda]\Pi^{\Lambda_\tau}_mU_{\tau,0}[\Lambda]\Pi^{\Lambda_0}_n\right) \delta( w, E^{\Lambda_\tau}_m - E^{\Lambda_0}_n) \nonumber\\
                                      &=&  \frac{\e^{\beta  w} }{Z(\Lambda_0)} \sum_{m,n}    \e^{-\beta E^{\Lambda_\tau}_m}  \tr  \left(U^\dagger_{\tau,0}[\Lambda]\Pi^{\Lambda_\tau}_mU_{\tau,0}[\Lambda]\Pi^{\Lambda_0}_n\right) \delta(- w, E^{\Lambda_0}_n - E^{\Lambda_\tau}_m) \nonumber \\
                                      &=&  \e^{\beta(  w- \Delta F)} \sum_{m,n}    \frac{\e^{-\beta E^{\tilde \Lambda_0}_m}}{Z(\tilde \Lambda_0)}  \tr  \left(U_{\tau,0}[\Lambda]\Pi^{\eta_Q\tilde \Lambda_\tau}_nU^\dagger_{\tau,0}[\Lambda]\Pi^{\eta_Q\tilde \Lambda_0}_m\right) \delta(- w, E^{\tilde \Lambda_\tau}_n - E^{\tilde \Lambda_0}_m), \nonumber\\
                                      \label{eqn:backwardprot}                                      
\end{eqnarray}
where 
we used the definition~\eqref{eq:backward protocol} of the backward protocol $\tilde \Lambda$, and the assumptions~\eqref{eq:timereveigen}.
Then, we use  microreversibility, namely \eqref{eqn:qmirev} for $t=0$ and  obtain that
\begin{equation}
U_{\tau,0}^\dagger[\Lambda]=\Theta U_{\tau,0}[\tilde \Lambda] \Theta
\end{equation}
and 
\begin{equation}
U_{\tau,0}[\Lambda]=\Theta U_{\tau,0}^\dagger[\tilde \Lambda] \Theta,
\end{equation}
therefore
\begin{eqnarray}
& &\frac{P[ w;\Lambda] }{\e^{\beta(  w- \Delta F)}}=   \sum_{m,n}    \frac{\e^{-\beta E^{\tilde \Lambda_0}_m}}{Z(\tilde \Lambda_0)}  \tr  \left(U_{\tau,0}[\Lambda]\Pi^{\eta_Q\tilde \Lambda_\tau}_nU^\dagger_{\tau,0}[\Lambda]\Pi^{\eta_Q\tilde \Lambda_0}_m\right) \delta(- w, E^{\tilde \Lambda_\tau}_n - E^{\tilde \Lambda_0}_m) \nonumber \\
                                      & & =    \sum_{m,n}    \frac{\e^{-\beta E^{\tilde \Lambda_0}_m}}{Z(\tilde \Lambda_0)}  \tr  \left(\Theta U_{\tau,0}^\dagger[\tilde \Lambda] \Theta \Pi^{\eta_Q\tilde \Lambda_\tau}_n\Theta U_{\tau,0}[\tilde \Lambda] \Theta \Pi^{\eta_Q\tilde \Lambda_0}_m\right) \delta(- w, E^{\tilde \Lambda_\tau}_n - E^{\tilde \Lambda_0}_m)\nonumber \\  
                                      & & =   \sum_{m,n}    \frac{\e^{-\beta E^{\tilde \Lambda_0}_m}}{Z(\tilde \Lambda_0)}  \tr  \left( U_{\tau,0}^\dagger[\tilde \Lambda]  \Pi^{\tilde \Lambda_\tau}_nU_{\tau,0}[\tilde \Lambda]  \Pi^{\tilde \Lambda_0}_m\right) \delta(- w, E^{\tilde \Lambda_\tau}_n - E^{\tilde \Lambda_0}_m)  \nonumber\\
                                      & & =    P[- w; \tilde\Lambda] ,  
\label{eqn:commproj}
\end{eqnarray}
where 
we used again~\eqref{eq:timereveigen}
and  $\Theta^2=\mathbb{I}$.

\section*{Acknowledgments}

We would like to thank the  organizers, Beppe Marmo, and Alberto Ibort for their kindness in inviting us and for the effort they exerted on the organization of the workshop. 
This work was supported by Cohesion and Development Fund 2007-2013 - APQ Research Puglia Region ``Regional program supporting smart specialization and social and environmental sustainability - FutureInResearch,'' by the Italian National Group of Mathematical Physics (GNFM- INdAM), and by Istituto Nazionale di Fisica Nucleare (INFN) through the project ``QUANTUM.''

\end{document}